\documentclass[preprint]{elsarticle}
\usepackage{amsmath}
\usepackage[final]{pdfpages}
\usepackage{graphicx,bm}
\begin{document}

\title{On the Dissipative Version of the Gross--Pitaevski Equation}
\author[cft]{ K. Paw\l owski \corref{cor1}}

\author[cft]{ \L. A.Turski}

\cortext[cor1]{Corresponding author}

\address[cft]{Center for Theoretical Physics, Polish Academy of
Sciences. Al.\ Lotnik\'{o}w 32/46, 02-668,
Warszawa, Poland}


\date{\today}

\begin{abstract}
We propose a novel version of the dissipative Gross--Pitaevski equation and examine its properties. In contrast to previous  proposals our approach,   based on the metriplectic formulation of the dissipative system dynamics, conserves the number of particles in the system.
\end{abstract}

\maketitle

The rush to describe Bose-Einstein condensates discovered in clouds of atoms in optical traps catapulted the semi-classical equation proposed to describe the properties of the superfluid helium  by Gross and Pitaevski \cite{G1,P1,PS1}, akin the Ginzburg-Landau equations of superconductivity, into one of the main equations of theoretical many-body physics. This non-linear equation is mathematically equivalent to the non-linear Schr\"odinger equation with the quartic non-linearity and is conservative in the sense that the Lyapunov functional for it (equivalent to the  Ginzburg-Landau free energy) is a constant of motion. The other relevant constant of motion for the G-P equation is the number of particles in the system defined as the normalization of the condensate wave -function.  Various attempts have been made recently to generalize that equation in order to describe the dissipation mechanisms within the scheme of the Gross-Pitaevski model of the condensate dynamics \cite{P2,B1,S1}. The physical discussion of the meaning of that kind of damping has been given particularly clear in \cite{B1}. Those models describe dissipation mechanism which do not conserve both number of particles and the value of the Lyapunov functional. Since any system loosing particles is non conservative (energy being extensive quantity decreases with decrease of number of particles) it is of at least some interest to question how the dissipation conserving the number of particles can be build in the Gross-Pitaevski formulation of the condensate dynamics. We propose here a way to do so following scheme of the so-called metriplectic dynamics developed quite a time ago for other purposes in many body physics: ferromagnetism, plasma physics: both non-relativistic and relativistic, classical hydro- and magnetohydrodynamics. The metriplectic formulation could also be used for description of damping in quantum mechanism and there it results in the so-called Gisin equation.
 We will show that it can also be used to derive dissipative Gross-Pitaevski equation which describes the condensate dynamics with conserved number of particles (normalization of the condensate wave function).
   We will also show how that equation can be used to describe the condensate dynamics.
   The review of the other methods describing dissipative dynamics of the Bose condensate is given in \cite{jackson2008}.
 
 The usual form of the Gross-Pitaevski equation for the complex condensate wave function $\psi(\mathbf{x},t)$ in case of short range interparticle interactions reads
 \begin{equation}
  \label{GP1}
i\hbar\partial_t\psi(\mathbf{x},t)= -\frac{\hbar^2}{2m}\nabla^2\psi(\mathbf{x},t) + (V\psi(\mathbf{x},t)-\mu)+g|\psi(\mathbf{x},t)|^2\psi(\mathbf{x},t)
\end{equation}
where $V$ stands for ``external potential'' and which explicit form is irrelevant for our purposes, an g is the coefficient related to the scattering length of the system inter-particle interactions.

 Eq.\eqref{GP1} can be rewritten using the Poisson brackets between canonically conjugated fields $\psi$ and $\psi^*$
 \begin{equation}
  \label{Poisson}
  \{\psi(\mathbf{x},t),\psi^*(\mathbf{x'},t)\}=-\frac{i}{\hbar}\delta(\mathbf{x}-\mathbf{x'})
\end{equation}
and the Lyapunov (or free energy) functional:
\begin{eqnarray}
  \label{FreeE1}
  \mathcal{F}[\psi,\psi^*]&=& \int d^d x \left(\frac{\hbar^2}{2m}|\nabla\psi(\mathbf{x},t)|^2 + f(\psi,\psi^{*})\right)\nonumber\\
  &=&\int d^{d}x H(\psi,\psi^{*}),
  \end{eqnarray}
  where
  \begin{eqnarray}
\label{local f}
f(\psi,\psi*)=V|\psi(\mathbf{x},t)|^2+\frac{1}{2}g|\psi(\mathbf{x},t)|^4
\end{eqnarray}

as:
\begin{equation}
  \label{GP2}
  \partial_{t} \psi(\mathbf{x},t))=\{\psi(\mathbf{x},t), \mathcal{F}\left[\psi,\psi^*\right]\}=\frac{1}{i\hbar}\frac{\delta \mathcal{F}\left[\psi,\psi^*\right]}{\delta \psi^*(\mathbf{x},t)}
\end{equation}

It is easy to see that the particle density, defined as $\rho(\mathbf{x},t)=|\psi(\mathbf{x},t)|^2$ obeys the continuity equation:
\begin{equation}
  \label{continuity}
  \partial_t\rho(\mathbf{x},t)=-\nabla\cdot\mathbf{J}(\mathbf{x},t)
\end{equation}
with $\mathbf{J}(\mathbf{x},t)=\frac{\hbar}{2m i}(\psi^*(\mathbf{x},t)\nabla \psi(\mathbf{x},t)-c.c)$ representing particle current.

Following Eq.(\ref{continuity}) the total number of particles $\int d^d x\rho=||\psi||^2= N$ is conserved.  The trivial consequence of the equation  Eq.(\ref{GP2}) and its counterpart for $\psi^*$ is that the free energy $\mathcal{F}$ is conserved: $i \hbar\partial_{t}\mathcal{F}=\{\mathcal{F},\mathcal{F}\}=0$.

The dissipation has been added to the Gross-Pitaevskii equation, following the suggestion of Pitaevskii \cite{P2}, by rewriting Eq.(\ref{GP2}) as:
\begin{eqnarray}
\label{GPD1}
 \partial_{t} \psi(\mathbf{x},t))=(1-i \lambda)\frac{1}{i\hbar}\frac{\delta \mathcal{F}\{\psi,\psi^*\}}{\delta \psi^*(\mathbf{x},t)}
\end{eqnarray}
where $\lambda$ is the dissipation coefficient. It is easy to check that the consequence of Eq.(\ref{GPD1}) is that the number of particles in the system $N$ is decreasing at the rate $\lambda$.

Applications of the Eq. (\ref{GPD1}) have been discussed extensively in the literature \cite{P2,B1}. It can be easily generalized by combining it into so-called hybrid Gross-Pitaevski-Boltzmann equation by supplement it with rate equation for  depletion of the condensate $dN/dt$ \cite{Hybrid}.

What we will do in the following is to generalize Eq. (\ref{GPD1})  following the metriplectic formulation of the many body dynamics \cite{LAT1,LAT2, LAT3}, that is be replacing the Poisson bracket Eq.(\ref{Poisson}) by the metriplectic one:
\begin{eqnarray}
\label{metriplectic}
\{\{\psi(\mathbf{x}),\psi^{*}(\mathbf{x'})\}\}&=&  \{\psi(\mathbf{x}),\psi^{*}(\mathbf{x'})\} \nonumber \\ &-&\frac{\lambda}{\hbar}\left(\delta(\mathbf{x}-\mathbf{x'})-\frac{\psi(\mathbf{x})\psi^{*}(\mathbf{x'})}{||\psi||^{2}}\right)
\end{eqnarray}
where $\lambda$ is the dissipation coefficient.  In the quantum mechanics that procedure leads to the  Gisin generalization of the Schr\"odinger equation \cite{Gisin}. The properties of that generalization have been also discussed in \cite{LATG1}.

The dissipative Gross--Pitaevskii equation we propose assumes now the form:
\begin{eqnarray}
\label{GPD2}
\partial_{t}\psi(\mathbf{x},t)&=&\{\{\psi(\mathbf{x},t), \mathcal{F}\}\}\nonumber \\
&=&(1-i\lambda)\frac{1}{i\hbar}\frac{\delta\mathcal{F}}{\delta\psi^{*}(\mathbf{x},t)} \nonumber\\&+&\frac{i\lambda}{||\psi||^{2}}\int d^{d}x' \psi(\mathbf{x})\psi^{*}(\mathbf{x'})\frac{1}{i\hbar}\frac{\delta\mathcal{F}}{\delta\psi^{*}(\mathbf{x'})}
\end{eqnarray}
This integro--differential equation can be rewritten into the form
\begin{eqnarray}
\label{NL}
i\hbar\partial_{t}\psi(\mathbf{x},t)=(I-i\lambda\hat{Q})\star\frac{\delta\mathcal{F}}{\delta\psi^{*}}=\mathcal{D}(\psi,\psi^{*})\star\frac{\delta\mathcal{F}}{\delta\psi^{*}}
\end{eqnarray}
where the kernel $\mathcal{D}$ reads
\begin{eqnarray}
\mathcal{D}(\psi,\psi^{*})=
\delta(\mathbf{x} -\mathbf{x'})-i\lambda Q\left[\psi(\mathbf{x}),\psi^{*}(\mathbf{x'})\right]
\end{eqnarray}
Here the kernel $Q(\psi, \psi^{*})$ serves as the projection operator onto the direction perpendicular to the condensate wave function $\psi$ and $\star$ denotes integration with respect to space coordinates. In the conventional quantum mechanical notation it can be written as $Q=1-|\psi\rangle\langle\psi|/\|\psi\|^{2}$.
 One of the common methods to study properties of the condensate Bose system is the imaginary time evolution method (ITE).
  Careful analysis of the ITE method shows that it corresponds to the time evolution described by Eq.(\ref{GPD2}) in the limit of the very short evolution time and when $\lambda \rightarrow \infty$.
  
Inspite of its integro--differential form  Eq.(\ref{GPD2}) still preserves the normalization of the condense wave function.
Using  Eq.(\ref{GPD2}) we can derive the continuity equation in the form:
\begin{eqnarray}
\label{continuity2}
\partial_{t}(|\psi(\mathbf{x},t)|^{2})= &-&(1-i\lambda)\nabla\cdot\mathbf{J}\nonumber \\ &-&\frac{i\lambda |\psi(\mathbf{x})^{2}}{\|\psi\|^{2}}\int d^{d} x'\nabla\cdot\mathbf{J}(\mathbf{x'})
\end{eqnarray}
Integrating Eq.(\ref{continuity2})  we find that the number of particles $N=||\psi||^{2 }$  is conserved, similarly as in the case of the continuity equation Eq.(\ref{continuity}). This fact is a direct consequence of the construction of the metriplectic bracket describing dissipation happening on the simplectic leaf, $\mathcal{C}[\psi]=const$, defined by the functional $\mathcal{C}[\psi]$ which is constant of motion,  independently of the form of the system Hamiltonian,  just as a consequence of the  Poisson brackets \cite{LATG1} . In our case $\mathcal{C}[\psi]$ is the condensate wave function norm $\|\psi\|^2$.

The next consequence of our construction of the dissipative equation Eq.(\ref{GPD2}) is that the free energy--Lyapunov  functional  for our model cannot increase in time.  Indeed
\begin{eqnarray}
\label{FEdissipation}
\frac{d\mathcal{F}}{dt}=\{\{F,F\}\|\leq 0\,.
\end{eqnarray}

 From Eq. (\ref{FEdissipation}) it follows also that the mean value of the "energy" $H(\psi,\psi^{*})$, namely  $\int d^{d}x\,H(\psi,\psi^{*})=\langle H \rangle$ decreases in time. Indeed defining the generalized chemical potential
 \begin{eqnarray}
\label{chemical potential}
\mu[\psi,\psi^{*}]=\frac{\delta \mathcal{F}(\psi, \psi^{*})}{\delta \psi(\mathbf{x})}
\end{eqnarray}
we  can rewrite Eq.(\ref{FEdissipation}) as
\begin{eqnarray}
\label{FEdissipation2}
\hbar\frac{d\langle H \rangle }{dt}= - \lambda\big ( \langle \mu^{2}\rangle -\langle \mu\rangle^{2}\big )
\end{eqnarray}

 The stationary solutions of the dissipative Gross-Pitaevski equation Eq.(\ref{GPD2}) are the extrema of the free energy functional Eq.(\ref{FreeE1}) $\delta\mathcal{F}/\delta\psi^{*}=0$. Among interesting stationary solutions of that equation are the finite amplitude periodic in space solutions $\propto \exp(i \mathbf{K}\cdot\mathbf{x})$ and in the one-dimensional space  the dark solitons  $\psi_{ds}(x)$.  It is worth noting that the solitons with non-zero speed, i.e. the gray solitons, are not extrema of the free energy functional Eq.(\ref{FreeE1}).
 
  Let $\Psi(\mathbf{x})$ be one of those solutions. Linearizing Eq.(\ref{GPD2}) $\psi=\Psi+\Delta \psi$ one  finds that the excess free energy due to those fluctuations $\Delta \mathcal{F}$ decrease towards the value of the system free energy corresponding to the stationary solution $\Psi$.
 \begin{eqnarray}
\label{decrease}
\frac{d}{dt}\Delta\mathcal{F}=-\frac{2\lambda}{\hbar}\frac{\delta\Delta\mathcal{F}}{\delta\psi^{*}}\star Q\star \frac{\delta\Delta\mathcal{F}}{\delta\psi}
\end{eqnarray}
 
 The same linearization procedure allows us to derive dispersion relation for the fluctuations $\Delta \psi$ which correspond to Bogolons obtained by conventional analysis of  Eq.(\ref{GP1}).
 The linearized equations for fluctuations $\Delta\psi, \Delta \psi^{*}$ read:
  \begin{eqnarray}
\label{LFE}
i\hbar\partial_{t}\Delta\psi(\mathbf{x},t)&=\int d^{d}x' \mathcal{D}(\Psi(\mathbf{x}),\Psi^{*}(\mathbf{x}'))\nonumber\\
&\left(\mathcal{K}\Delta\psi(\mathbf{x}')+g\Psi^{2}\Delta\psi^{*}(\mathbf{x}')\right)\nonumber\\
i\hbar\partial_{t}\Delta\psi^{*}(\mathbf{x},t)&=\int d^{d}x' \mathcal{D}^{*}(\Psi(\mathbf{x}),\Psi^{*}(\mathbf{x}'))\nonumber\\
&\left(\mathcal{K}\Delta\psi^{*}(\mathbf{x}')+g{\Psi^{*}}^{2}\Delta\psi(\mathbf{x}')\right)
\end{eqnarray}
where $\mathcal{K}=(-\hbar^{2}\nabla^{2}/2m +(V-\mu)+ 2g|\Psi|^{2})$.
The integral term on the RHS of Eqs.(\ref{LFE}) guarantee that the condensate is stable with respect to $k\rightarrow 0$ perturbations.
For real stationary solutions $\Psi$ we can rewrite  Eqs.(\ref{LFE}) introducing the real and imaginary parts for $\Delta\psi$: $\Delta\psi+\Delta\psi^{*}=2\chi,\Delta\psi-\Delta\psi^{*}=2i\eta$.
  Assuming now that the functions $\chi,\eta \propto \exp(\pm i\omega t+\imath \mathbf{k}\cdot\mathbf{x})$  and for homogeneous condensate solution $\Psi, |\Psi|^{2}=n_{0}$ ( case without trapping potential $V=0$ ) the integrals on the RHS of Eqs \eqref{LFE} contribute terms proportional to $\delta_{\mathbf{k},0}$ which cancels with proper terms on the LHS of the same equations.  Eqs (\ref{LFE}) reduce then to  the pair of algebraic equations for $\chi$ and $\eta$ resulting in the dispersion relation $\omega(\mathbf{k})$
  which for small damping constant $\lambda$ and long wavelengths reduces to the damped sound waves dispersion relation
 \begin{eqnarray}
\label{sound}
\omega\approx c_{s}k-i \lambda\frac{\hbar\mathbf{k}^{2}}{2m}
\end{eqnarray}
with $c_{s}=\hbar\sqrt{n_{0}g/m}\,$ -- the sound velocity in the condensate.

\begin{center}
\begin{figure}
 \includegraphics[width=9.5cm]{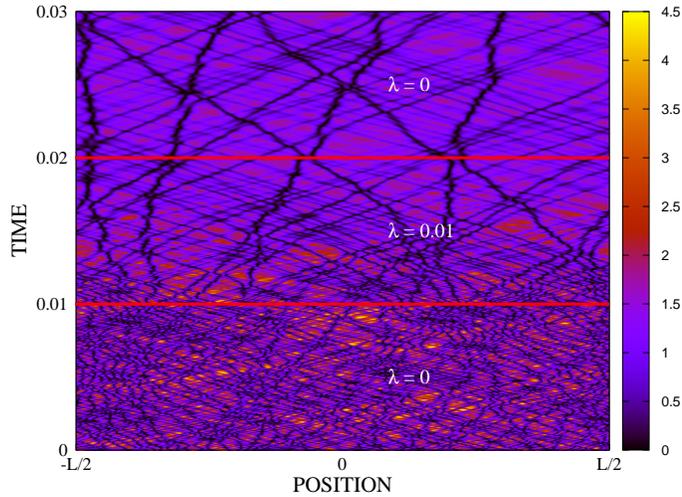}
 \caption{The density of a gas as a function of time and position. The evolution is divided into three stages, during which the dissipation parameters is either $0$ or $0.01$ as marked on the graph.Parameters: $gN = 10^4$, initial state comes from the thermal ensemble with the temperature $T=7\times 10^4$. \label{fig:KB}}
 \end{figure}
 \end{center}

We close our analysis with application of Eq. \eqref{GPD1}  to describe property of the slowly cooled Bose system.
To do so we investigated two cases.
The first one is the cooled Bose gas which should show reminiscence of the Kibble-Zurek  mechanism of the domain formation. The domain walls created in such a mechanism are usually identify with the dark solitons \cite{WDGR}. The second case is the time evolution of the cooled Bose system with initial conditions corresponding to existence in it two dark soliton $\psi_{s}(x)$ , i.e stationary solutions of  $\delta\mathcal{F}/\delta\psi^{*}=0$. In both simulations  our system is placed in a box of dimension $L$ with periodic boundary conditions and we scale our variables  using $L, mL^{2}/\hbar, \hbar^{2}/mL^{2}$ and $\hbar^{2}/mL^{2}k_{B}$  for length, time, energy and temperature,respectively. Here $k_{B}$ is the Boltzmann constant.

\begin{center}
\begin{figure}
	\begin{minipage}{0.45\textwidth}
		\centering
		
		(a)
		
		 \includegraphics[width=\textwidth]{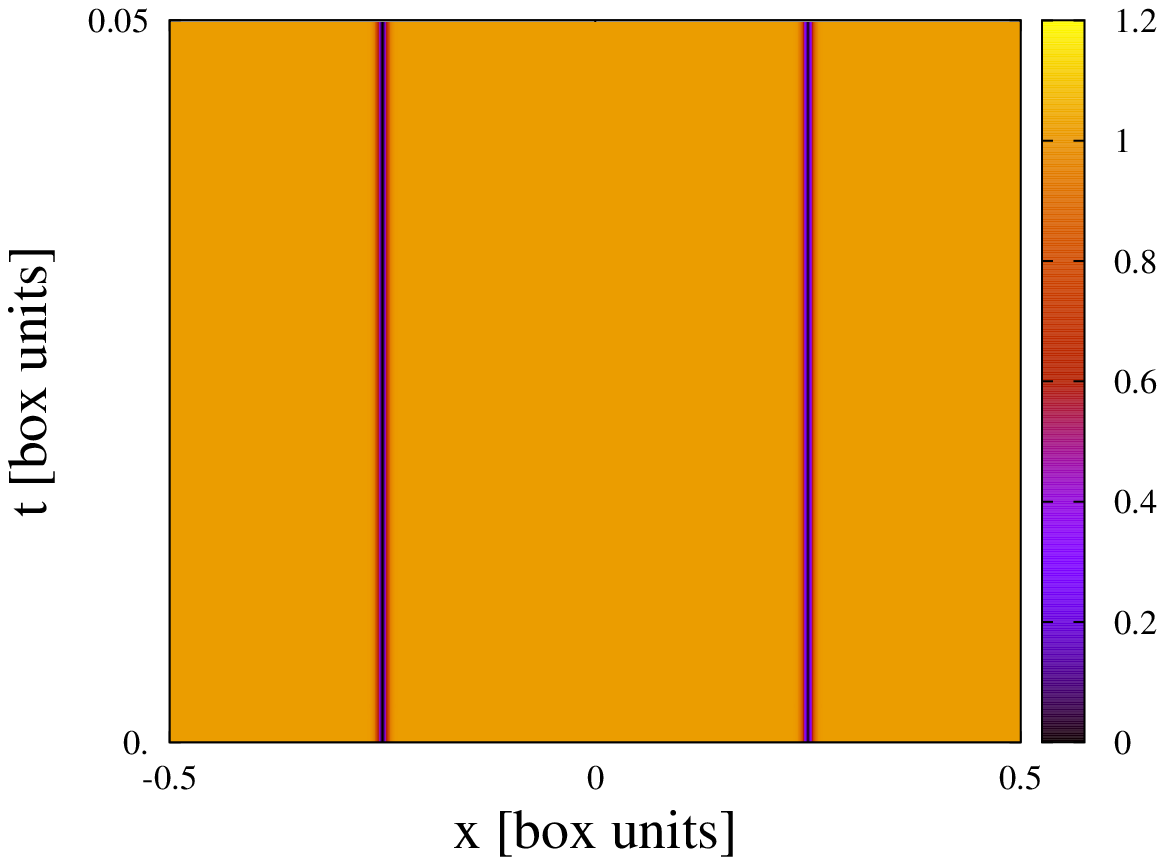}
	\end{minipage}\hspace{0.03\textwidth}
	\begin{minipage}{0.45\textwidth}
		\centering
		
		(b)
		
		\includegraphics[width=\textwidth]{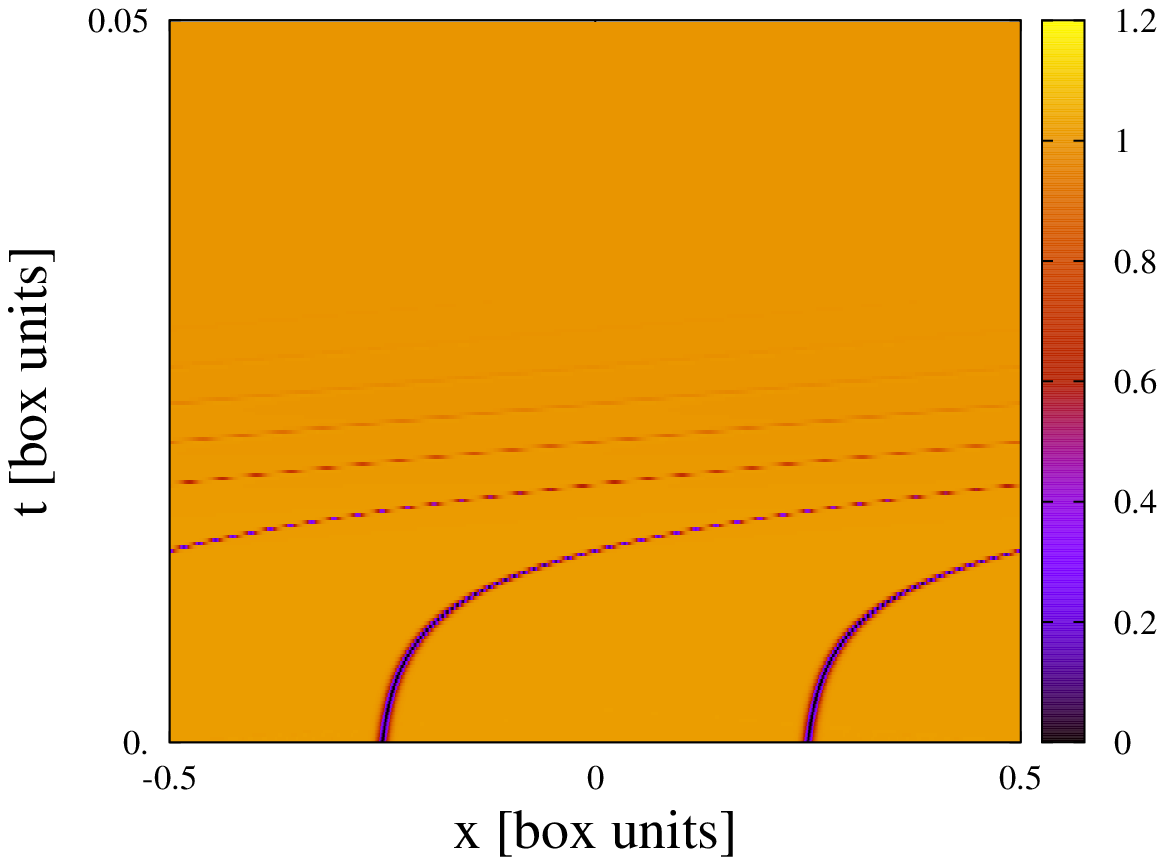}
	\end{minipage}\hspace{0.03\textwidth}
 \caption{The dynamics of density of a gas in one dimensional box with periodic boundary conditions as a function of time and position.  The initial state consists of two solitons \cite{ShabatandZakharov} placed at $x=\pm 0.2 L$ with the same speeds equal to (a) $v=0$ (black solitons) and (b) $v=0.01$ of speed of sound (slow gray solitons). 
 Parameters: $gN=4\times 10^4$, $\lambda=0.01$, position and time are in the box units, $L$ and $(m L^2)/\hbar$, where $L$ is the length of the box.
 \label{fig:2sol}}
 \end{figure}
 \end{center}
 
Our first simulation runs as follow:
\begin{enumerate}
    \item  Following \cite{WGR} we generate an initial state, that is  field $\psi(x, t=0)$, from a thermal ensemble with a high temperature $T=7\times 10^4$ in a box units. The method \cite{WGR} uses the classical field approximation, in which a thermal ensemble is represented by a collection of many fields generated via Monte-Carlo method.
    \item Next we  propagate the field $\psi(x, t=0)$ for a time $T_{\rm TH}$ using Eq.(\ref{GPD2}) with, $\lambda = 0$. This is the technical step during which we monitor the changes of the occupation of the ground state, $\psi_{\rm GS} = 1/\sqrt{L}$ and verify that it has the expected fluctuation and hence it is a good thermal sample.
    \item We abruptly switch on the dissipation $\lambda = 0.01$ and we let the system evolve for a the time $T_{\rm DIS} = 0.01$ (box units).
    \item We switch the dissipation off, setting $\lambda$ to $0$, and further evolve the field during the time $T_{\rm SOL}$.
\end{enumerate}
An example of such evolution is given in Fig. \ref{fig:KB}.

Our second simulation uses Eq.(\ref{GPD2}) to propagate initial conditions of the cooled Bose gas consisting of two solitons.  Fig. \ref{fig:2sol} shows our results. 
 As shown in the panel (a) the damping does not affect the black solitons $\psi_{ds}$, being 
an	extremum  of the free energy functional $\mathcal{F}[\psi,\psi^*]$.
On the other hand the gray solitons $\psi_{gs}$ are not such extrema, i.e. $\frac{\delta \mathcal{F}[\psi,\psi^*]}{\delta \psi^*}\mid_{\psi_{gs}}\neq 0$ and even if they have an infinitesimal speed they will eventually decay. An example is given in  Fig. \ref{fig:2sol}b).
 Due to dissipation the gray solitons are getting shallower and accelerate reaching finally a stable state.

We have presented dissipative generalization of the Gross-Pitaevsky equation  (\ref{GPD2}) which preserves the number of particles in the condensate. We have analyzed properties of this equation, show how damping affects the Bogolyubov excitations and illustrate properties of our description of damping on two numerical examples of the  cooled Bose gas behavior.
 
\textbf{Acknowledgment}
We  would like to thank  Maciej Lewenstein for discussions.
K. P. acknowledges support by the (Polish) National Science Center Grant No. 2014/13/D/ST2/01883.


\end{document}